\documentclass[11pt]{article}
\usepackage[utf8]{inputenc}
\usepackage[T1]{fontenc}
\usepackage{amsmath,amssymb,amsfonts}
\usepackage{graphicx}
\usepackage{geometry}
\usepackage{authblk}
\usepackage{hyperref}
\usepackage{booktabs}
\usepackage{siunitx}
\usepackage[numbers,sort&compress]{natbib}
\usepackage{doi}
\hypersetup{colorlinks=true, linkcolor=blue, citecolor=blue, urlcolor=blue}

\title{Kuramoto Phase Synchronization in Regional Epidemic Dynamics:\\
Two Test Cases from European COVID-19 and Influenza Surveillance}

\author[1]{D. Delepine}
\author[2]{J. de J. Bernal-Alvarado}
\affil[1]{Departamento de F\'isica, Divisi\'on de Ciencias e Ingenier\'ias, Universidad de Guanajuato, M\'exico}
\affil[2]{Departamento de Ingenier\'ia F\'isica, Divisi\'on de Ciencias e Ingenier\'ias, Universidad de Guanajuato, M\'exico}

\date{\today}

\begin{document}
\maketitle

\begin{abstract}

We test whether the Kuramoto model of coupled phase oscillators provides a quantitative description of spatial synchronization in regional epidemic waves, using two independent surveillance datasets: daily COVID-19 incidence for the 400 German \emph{Kreise} (districts), 2021--2023, and weekly Influenza-Like-Illness (ILI) consultation rates for 12 European countries with sufficiently complete European Centre for Disease Prevention and Control (ECDC) surveillance records, 2021--2026. In both cases, after isolating each unit's oscillatory component via bandpass filtering and Hilbert-transform phase extraction, we find a high global Kuramoto order parameter (German districts: $\langle r\rangle \approx 0.86$--$0.88$; European countries: $\langle r\rangle \approx 0.97$), indicating strong synchronization at the respective national and continental scale. In both cases we show, however, that this global synchronization is dominated by a common-mode driver rather than by pairwise local coupling. Removing the common mode and examining the \emph{residual} phase dynamics reveals genuine short-range synchronization in the German data: the excess local order parameter decaying exponentially with distance and a fitted correlation length $\xi = 152 \pm 6$~km. The same qualitative pattern — significantly higher pairwise phase-locking between geographic neighbors than non-neighbors (Mann--Whitney $p=0.005$) and a residual local-synchronization signal peaking at $R\approx 700$--$1500$~km — appears in the European country-level panel, though with much lower statistical power given the small number (12) of countries with sufficiently complete data. We interpret the convergent pattern in two independent diseases, spatial scales, and geographic units as evidence for a general two-scale structure in regional epidemic synchronization: a long-range, high-amplitude common-mode field superimposed on a shorter-range, genuinely diffusive local coupling of the kind the Kuramoto model was designed to describe. Finally, we cross-validate the German phase-synchronization observables against a companion gauge-mediated, Doi--Peliti field-theoretic framework fit to the same incidence data: the district-level effective screening mass $m_R$ of that framework is found to be empirically uncorrelated with the local Kuramoto order parameter (a clean null result), while the national-aggregate $(m_R, r)$ phase portrait independently reproduces a previously reported hysteresis loop, and the empirical pairwise phase lags quantitatively match, and structurally track the two asymptotic regimes of, a separate fractional-calculus prediction for the phase of the effective reproduction number. We test three candidate mechanisms for this hysteresis: a purely deterministic tree-level model is shown, on structural grounds, to be unable to reproduce it; a multi-wave sweep-rate test is inconclusive at the available sample size; and a network-level test finds a significant frequency-degree correlation, the structural signature of explosive synchronization. Searching for the real-world driver of this local coupling structure, we find a strong, distance-independent excess synchronization between districts sharing the same federal state (\emph{Bundesland}), consistent with shared school-holiday calendars and state-level pandemic policy as candidate mechanisms.
\end{abstract}

\section{Introduction}
\label{sec:intro}

The Kuramoto model \cite{kuramoto1975,strogatz2000} — a population of phase oscillators $\theta_i(t)$ with natural frequencies $\omega_i$ coupled through a sinusoidal interaction,
\begin{equation}
\dot\theta_i = \omega_i + \frac{K}{N}\sum_{j=1}^N A_{ij}\sin(\theta_j-\theta_i),
\label{eq:kuramoto}
\end{equation}
where $\theta_i(t)$ is the instantaneous phase of oscillator $i$, $\omega_i$ its natural frequency drawn from a distribution $g(\omega)$, $K\geq 0$ the global coupling strength, $N$ the total number of oscillators, and $A_{ij}\in\{0,1\}$ the entries of the network adjacency matrix ($A_{ij}=1$ if units $i$ and $j$ are coupled, $A_{ij}=0$ otherwise),
is the canonical minimal model for spontaneous synchronization in networks of coupled dynamical units, with applications ranging from neural oscillators to power grids. The collective synchronization state of the population is measured by the 
global Kuramoto order parameter $r(t)\in[0,1]$, defined via the mean-field 
complex amplitude
\begin{equation}
r(t)\,e^{i\psi(t)} = \frac{1}{N}\sum_{j=1}^{N} e^{i\theta_j(t)},
\label{eq:orderparameter}
\end{equation}
where $\psi(t)$ is the instantaneous mean-field phase. $r=0$ corresponds to 
complete incoherence (phases uniformly distributed on the circle) and $r=1$ 
to perfect synchronization (all phases identical).

Seasonal epidemics can be also considered as oscillators: incidence in a given region rises and falls with a roughly periodic structure set by depletion of susceptibles, seasonal forcing or successive variant waves. Whether this analogy is more than qualitative is the question this paper investigates  using two independent test cases chosen to differ as much as possible in disease, spatial scale, temporal resolution, and geographic unit while both remaining, in principle, within reach of routine public health surveillance data.

In our first study case, we use daily COVID-19 case counts across Germany's 400 \emph{Kreise} (districts) — a fine-grained, sub-national, daily-resolution dataset spanning several distinct pandemic waves. In the second study case,   weekly influenza-like-illness consultation rates across a panel of European countries reported to the European Centre for Disease Prevention and Control (ECDC) — a coarser-grained, international, weekly-resolution dataset dominated by a single strong annual seasonal cycle. If a qualitatively similar synchronization structure emerges independently in both cases that convergence is itself evidence that the structure reflects a general rule about spatially extended epidemic synchronization rather than an artifact specific to one dataset or disease.

In both cases our approach proceeds in the same stages: we extract an instantaneous phase $\theta_i(t)$ per unit via bandpass filtering and the Hilbert transform and compute the global Kuramoto order parameter then we test whether pairwise phase-locking is structured by geography; and finally, motivated by the results of that test, we decompose the dynamics into a common-mode field and a geographically local residual, characterizing the residual's spatial correlation structure directly. Case Study I additionally includes a fully data-driven attempt to reconstruct the Kuramoto coupling matrix $K_{ij}$ without assuming a geographic ansatz.

This paper brings together two largely separate literatures — spatial synchrony in infectious disease dynamics, and the Kuramoto/coupled-oscillator formalism.
\paragraph{Spatial synchrony in infectious disease dynamics}
The empirical observation that epidemics in neighboring regions rise and fall in a correlated, phase-locked manner has a long history in mathematical epidemiology. The best-studied lineage concerns pre-vaccination measles in England and Wales: \cite{grenfell2001} showed, using detailed weekly case records across cities and towns, that measles spread as a hierarchy of traveling waves radiating outward from large urban centers, with synchrony decaying with distance and with population size setting the local persistence threshold. The analysis in that work employed nonlinear time-series methods and mechanistic SEIR-type models; its central finding of a distance-decaying spatial correlation structure with a characteristic scale is directly analogous to the local-residual correlation length $\xi = 152 \pm 6$~km we report here for COVID-19.

The role of human mobility in driving epidemic synchrony was put on a quantitative footing for influenza by \cite{viboud2006}, who analyzed the spatiotemporal spread of seasonal influenza across the United States and showed that a gravity model of population-weighted mobility explained the synchrony structure  better than simple geographic distance, with large metropolitan areas serving as hubs from which synchrony radiated hierarchically. This is the direct predecessor of the gravity-model coupling $K^{\text{grav}}_{ij}$ we test in Sec.~\ref{sec:case1-methods}, and the dominance of a long-range common-mode field over local diffusion that we find here is consistent with the hierarchical picture of \cite{viboud2006}, in which national-scale mobility creates near-simultaneous seeding of many localities rather than a sequentially diffusing front.

At a global scale, \cite{colizza2006} demonstrated that the worldwide airline transportation network provides the backbone for the spatial spread of pandemic pathogens, and that epidemic predictability at the global level is fundamentally constrained by the heterogeneity of that network. That line of work uses empirically measured mobility networks as a fixed, externally-imposed coupling structure, rather than attempting to infer the coupling structure from the epidemic phase data itself, as we do in Sec.~\ref{sec:case1-results-inverse}. % The identifiability problem we document there — that near-total synchronization renders the inverse problem ill-conditioned — is in this sense complementary to the forward-modeling approach of \citet{colizza2006}: when the coupling is strong enough to produce the high global order parameters we observe ($\langle r\rangle \approx 0.88$--$0.97$), it cannot be individually resolved from the resulting phase dynamics.

In the ecological context, \cite{blasius1999} showed that spatially extended systems of coupled chaotic oscillators — used as a model for ecological metapopulations — exhibit robust phase synchronization even when the amplitude dynamics remain chaotic and spatially heterogeneous, a phenomenon directly related to the common-mode / local-residual decomposition we employ: the global Kuramoto order parameter captures phase alignment independently of amplitude fluctuations. The related {\it Moran effect} — synchronization of spatially separated populations by a shared environmental driver \cite{moran1953} — is a natural null model for the common-mode field we identify and remove in Secs.~\ref{sec:case1-results-local} and \ref{sec:case2-results}; in our setting the analog of the Moran driver is the shared national (or continental) seasonality and mobility field.

The phase-locking value (PLV) we use to quantify pairwise synchronization (Sec.~\ref{sec:case1-methods}) was introduced in the neuroscience context by \cite{lachaux1999} as a measure of instantaneous phase coherence between simultaneously recorded brain signals.

 A series of recent descriptive analyses \cite{yoshikura2022a, yoshikura2022b} have noted synchronization of COVID-19 epidemic curves between neighboring countries using cross-correlation methods, but without phase extraction, order-parameter quantification, or common-mode decomposition. The present work thus provides the first quantitative Kuramoto-framework characterization of epidemic phase synchronization and its spatial structure.

\paragraph{The Kuramoto model and synchronization transitions.}The Kuramoto model \cite{kuramoto1975} and its subsequent mathematical analysis \cite{strogatz2000, acebron2005} constitute the canonical formalism for spontaneous synchronization in networks of coupled oscillators. \cite{rodrigues2016} provide a comprehensive review of Kuramoto dynamics on complex networks, covering the bifurcation structure, finite-size effects, and the role of network topology in shaping the onset and character of synchronization. % We draw directly on this body of theory in interpreting the relationship between the inferred coupling network and the observed synchronization structure (Secs.~\ref{sec:case1-results-inverse} and~\ref{sec:crossval-mechanisms}).

The most directly relevant theoretical result for our purposes is the phenomenon of \emph{explosive synchronization}: \cite{gomezgardenes2011,kuehn2021} showed that in heterogeneous oscillator networks where the natural frequency of a node is positively correlated with its degree, the onset of synchronization becomes discontinuous and hysteretic — a first-order transition rather than the second-order transition of the classical Kuramoto model. This is the mechanism we test in Sec.~\ref{sec:crossval-mechanisms} as a candidate explanation for the hysteresis loops observed in the $(m_R, r)$ phase plane.

The paper is organized in the following way. Section II are dedicated to COVID-19 analysis. The section is divided in three subsections: data and preprocessing, methods and results. In this section, the general framework of synchronization as a universal phenomenon in nonlinear science, including the Hilbert-transform approach to instantaneous phase extraction is developed. Section III corresponds to the study of the Study case II. In section IV, we compare our results to previous studies on spatial distribution of COVID-19 in Germany. In section V, we are looking for the structural cause of the explosive synchronization observed in previous section. Conclusions are presented in section VI.

\section{Study case I: COVID-19 Across German Districts}
\label{sec:case1}

\subsection{Data and preprocessing}
\label{sec:case1-data}

\subsubsection{Incidence data and district alignment}
Daily case counts by district and report date were obtained from the RKI's public COVID-19 line-list (\texttt{RKI\_COVID19}), aggregated into a district $\times$ day incidence matrix spanning 2 March 2020 to 22 January 2023 (1057 days). The raw aggregation contained 413 district columns which differs from the official number of 400 districts. This is  due to two artifacts of administrative history:
\begin{enumerate}
    \item  Berlin's twelve constituent boroughs (AGS 11001--11012) were reported individually \emph{in addition to} the standard district-level aggregate (AGS 11000)
    \item the city of Eisenach (AGS 16056) was reported separately for a short period preceding its 1 July 2021 administrative merger into the Wartburgkreis (AGS 16063)
\end{enumerate}
We resolved both cases in the following way:  folding the twelve Berlin boroughs into AGS 11000  and merging Eisenach's small residual case count into the Wartburgkreis series — yielding a clean $1057 \times 400$ matrix.

\subsubsection{Phase extraction}
For each district we computed incidence per 100{,}000 population, applied a 7-day centered moving average to remove weekday reporting artifacts, and took $\log(1+x)$ to stabilize variance across the wide dynamic range spanned by successive waves. We then applied a third-order Butterworth bandpass filter (passband: periods 
of 40--220 days, chosen to span the range of observed inter-wave intervals) 
and extracted the instantaneous phase via the Hilbert transform, 
$\theta_i(t) = \arg\left[\mathcal{H}\{x_i(t)\}\right]$, where 
$\mathcal{H}\{x_i(t)\} = x_i(t) + i\,\tilde{x}_i(t)$ and 
$\tilde{x}_i(t) = \frac{1}{\pi}\,\mathrm{P.V.}\!\int_{-\infty}^{\infty} 
\frac{x_i(\tau)}{t-\tau}\,d\tau$.

 Restricting the filter to the 
2021--2023 analysis window produces a  collapse of the order parameter 
at the window border ($r\approx 0.24$) because \texttt{filtfilt} lacks real data 
near the boundary to resolve the signal low-frequency content. The filter over the full available series from March 2020, is used but the data of year 2020 were only used as a warm-up buffer before restricting to the analysis 
window; this recovers a boundary value consistent with its neighbors 
($r\approx 0.84$). No equivalent correction is possible at the end of the 
series.  We therefore flagged the terminal 
$\sim$110 days (half the maximum filter period) as a zone of 
reduced reliability and were excluded them from some analyses, following standard 
practice in windowed spectral estimation \cite{torrence1998}.

\subsubsection{Geographic data}
District boundaries were obtained from  the official BKG \texttt{VG250} administrative boundary shapefile (land-area feature class, \texttt{GF=4}), which agrees with our 400-district AGS ordering. From this geometry we constructed (a) a binary Queen-contiguity adjacency matrix $A_{ij}$ (shared border or vertex; mean degree 5.22, no isolated districts), (b) the inter-centroid distance matrix $D_{ij}$ (computed in the equal-area projection EPSG:3035/UTM), and (c) a gravity-model coupling $K^{\text{grav}}_{ij} \propto (\text{pop}_i\,\text{pop}_j)^{a}/D_{ij}^{b}$ \cite{xia2004}.

\subsection{Methods}
\label{sec:case1-methods}

\subsubsection{Global order parameter}
The global Kuramoto order parameter was computed as $r(t)e^{i\psi(t)} = \frac{1}{N}\sum_{j=1}^{N} e^{i\theta_j(t)}$ over all 400 districts.

\subsubsection{Pairwise phase-locking value}
For each district pair we computed the phase-locking value $\text{PLV}_{ij} = \left|\left\langle e^{i(\theta_i(t)-\theta_j(t))}\right\rangle_t\right|$, averaged over the reliable portion of the analysis window (excluding the terminal cone of influence, leaving 641 days). The phase-locking value (PLV) between districts $i$ and $j$ is defined as
\begin{equation}
    \text{PLV}_{ij} 
    = \left|\left\langle e^{i\left(\theta_i(t)-\theta_j(t)\right)}\right\rangle_t\right|
    = \left|\frac{1}{T}\sum_{t=1}^{T} e^{i\left(\theta_i(t)-\theta_j(t)\right)}\right|,
\end{equation}
where $\theta_i(t)$ is the instantaneous phase of district $i$ at time $t$, 
$\langle\cdot\rangle_t$ denotes averaging over the analysis window of $T$ 
time steps.%Their 
%time-average is a vector whose length measures how consistently that 
%difference clusters around a fixed value.
$\text{PLV}_{ij}=1$ indicates 
perfect phase locking (constant phase difference), while 
$\text{PLV}_{ij}=0$ indicates a uniformly distributed phase difference, 
i.e.\ no synchrony.

We tested three geographic predictors of $\text{PLV}_{ij}$ via ordinary least squares on the Fisher-$z$-transformed PLV: binary adjacency, log-distance, and log-gravity (both with the canonical exponents $a=0.5,\,b=1$ and with $a,b$ fit freely by OLS).

\subsubsection{Data-driven coupling reconstruction}
In the weak-coupling regime, Eq.~\eqref{eq:kuramoto} is linear in the unknown couplings $K_{ij}$ for a given phase trajectory. For each district $i$ we estimated $\dot\theta_i$ by centered finite differences of the unwrapped phase and fit
\begin{equation}
\dot\theta_i(t) = \omega_i + \sum_{j\neq i} K_{ij}\sin\!\big(\theta_j(t)-\theta_i(t)\big) + \varepsilon_i(t)
\end{equation}
via $\ell_1$-regularized (Lasso) linear regression \cite{tibshirani1996}, with the regularization strength selected by 5-fold cross-validation on a representative subset of nodes and applied uniformly. This yields 400 independent sparse regressions (399 candidate predictors, 641 time points each).

\subsubsection{Common-mode removal and local order parameter}
Given a negative result at the network-reconstruction stage (Sec.~\ref{sec:case1-results-inverse}), we defined the residual phase $\tilde\theta_i(t) = \arg\!\big[e^{i(\theta_i(t)-\psi(t))}\big]$, i.e.\ each district's phase relative to the instantaneous national mean-field phase $\psi(t)$. We then defined, for each district $i$ and radius $R$, a \emph{local} order parameter restricted to the geographic neighborhood $\mathcal{N}_R(i) = \{j : D_{ij}\le R\}\cup\{i\}$,
\begin{equation}
r^{\text{loc}}_i(t; R) = \left|\frac{1}{|\mathcal{N}_R(i)|}\sum_{j\in\mathcal{N}_R(i)} e^{i\tilde\theta_j(t)}\right|,
\end{equation}
and its time- and district-averaged value $\langle r^{\text{loc}}(R)\rangle$. Statistical significance was assessed against a null distribution built from degree-matched random neighborhoods (same $|\mathcal{N}_R(i)|$, drawn uniformly at random rather than by geographic proximity), and a correlation length $\xi$ was extracted by fitting the excess $\langle r^{\text{loc}}(R)\rangle - \langle r^{\text{loc}}_{\text{null}}(R)\rangle \propto e^{-R/\xi}$. This same local-order-parameter machinery is reapplied without modification to Case Study II (Sec.~\ref{sec:case2-results}).

\subsection{Results}
\label{sec:case1-results}

\subsubsection{ Temporal structure of Kuramoto synchronization}
\label{sec:case1-results-global}

\begin{figure}[htbp]
\centering
\includegraphics[width=0.85\textwidth]{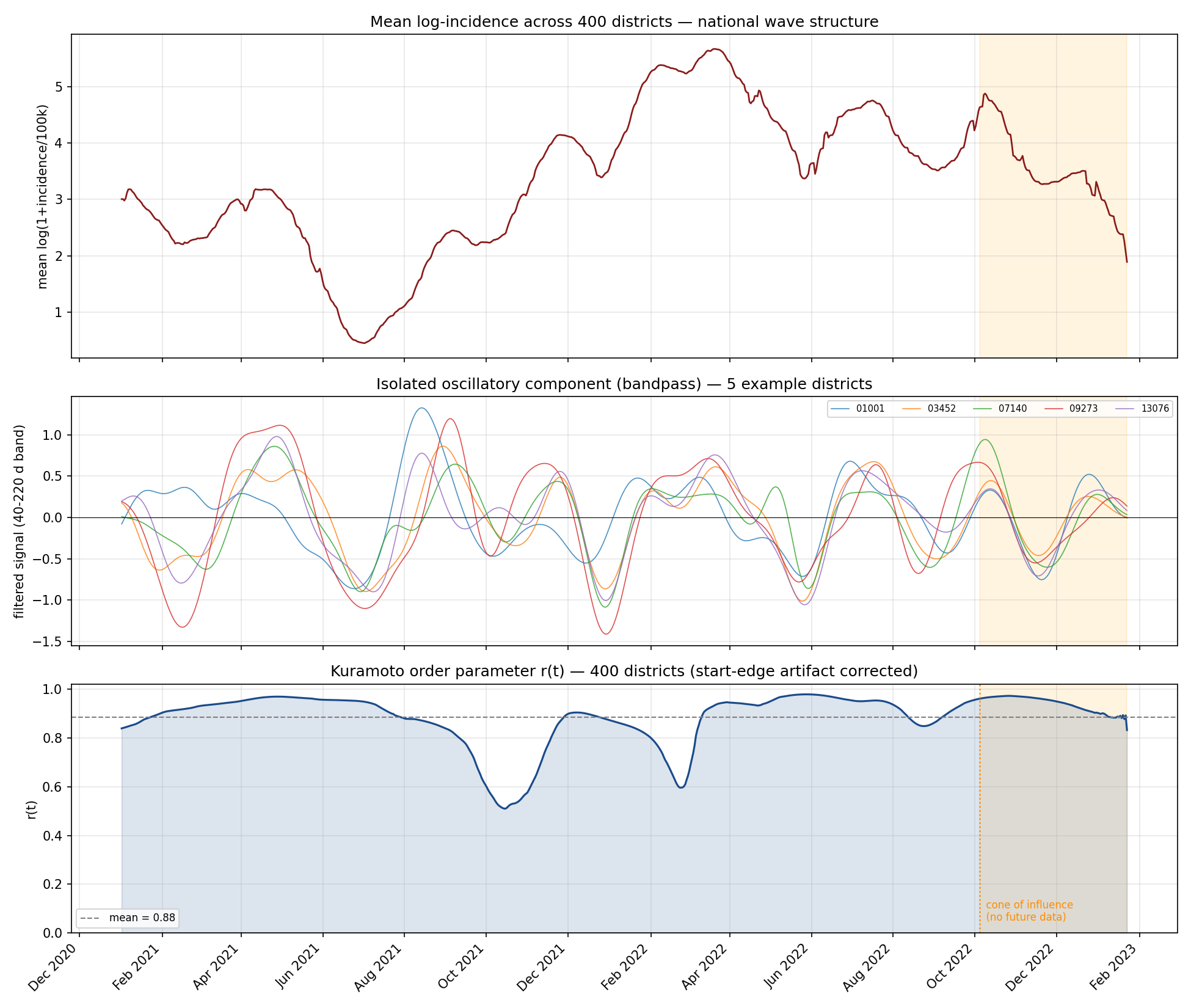}
\caption{Case Study I. Top: mean log-incidence per 100{,}000 across all 400 districts, showing the successive national waves. Middle: bandpass-filtered oscillatory component for five representative districts. Bottom: global Kuramoto order parameter $r(t)$; the shaded region marks the terminal cone of influence.}
\label{fig:order_parameter}
\end{figure}

 The  oscillatory component for a sample of districts alongside the global order parameter $r(t)$ is shown in Figure~\ref{fig:order_parameter}. The order parameter is given as $\langle r\rangle \approx 0.86$--$0.88$, and it appears to leading order as a  national phenomenon. Two statistically robust  episodes are nonetheless apparent, $r(t)$ falling in September--November 2021 and January--March 2022; both coincide with known epidemiologically heterogeneous periods (the Delta-to-pre-Omicron  and the BA.1-to-BA.2 transition, respectively)  supporting their interpretation as genuine partial desynchronization.

\subsubsection{Geographic structure of pairwise synchronization}
\label{sec:case1-results-plv}

\begin{figure}[htbp]
\centering
\includegraphics[width=0.95\textwidth]{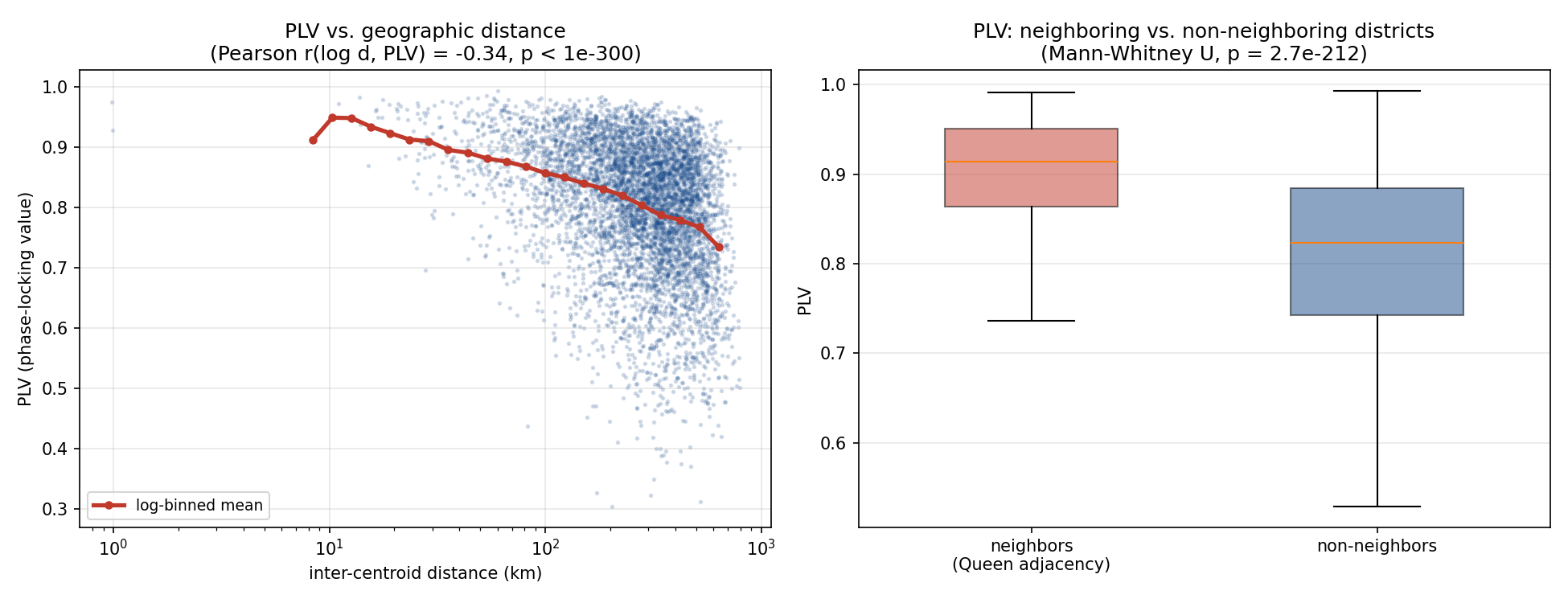}
\caption{Case Study I. Left: pairwise phase-locking value versus inter-centroid distance (light points: random subsample of district pairs; red curve: log-binned mean). Right: PLV distribution for geographic neighbors (Queen adjacency) versus non-neighbors.}
\label{fig:plv_geography}
\end{figure}

PLV is higher between geographic neighbors than non-neighbors (mean 
$0.896$ vs.\ $0.803$, $p = 2.7\times10^{-212}$) and declines with distance 
($r = -0.34$; Fig.~\ref{fig:plv_geography}), confirming that synchronization 
has genuine geographic structure. Geography nonetheless explains only a modest 
fraction of the variation in PLV: log-distance alone accounts for $R^2=0.143$, 
and adding population via a gravity model raises this only marginally to 
$R^2=0.146$. The fitted population exponent ($\alpha=0.019$) is an 
order of magnitude smaller than the canonical value of $0.5$.

\subsubsection{Data-driven coupling reconstruction: a cautionary result}
\label{sec:case1-results-inverse}

\begin{figure}[htbp]
\centering
\includegraphics[width=0.85\textwidth]{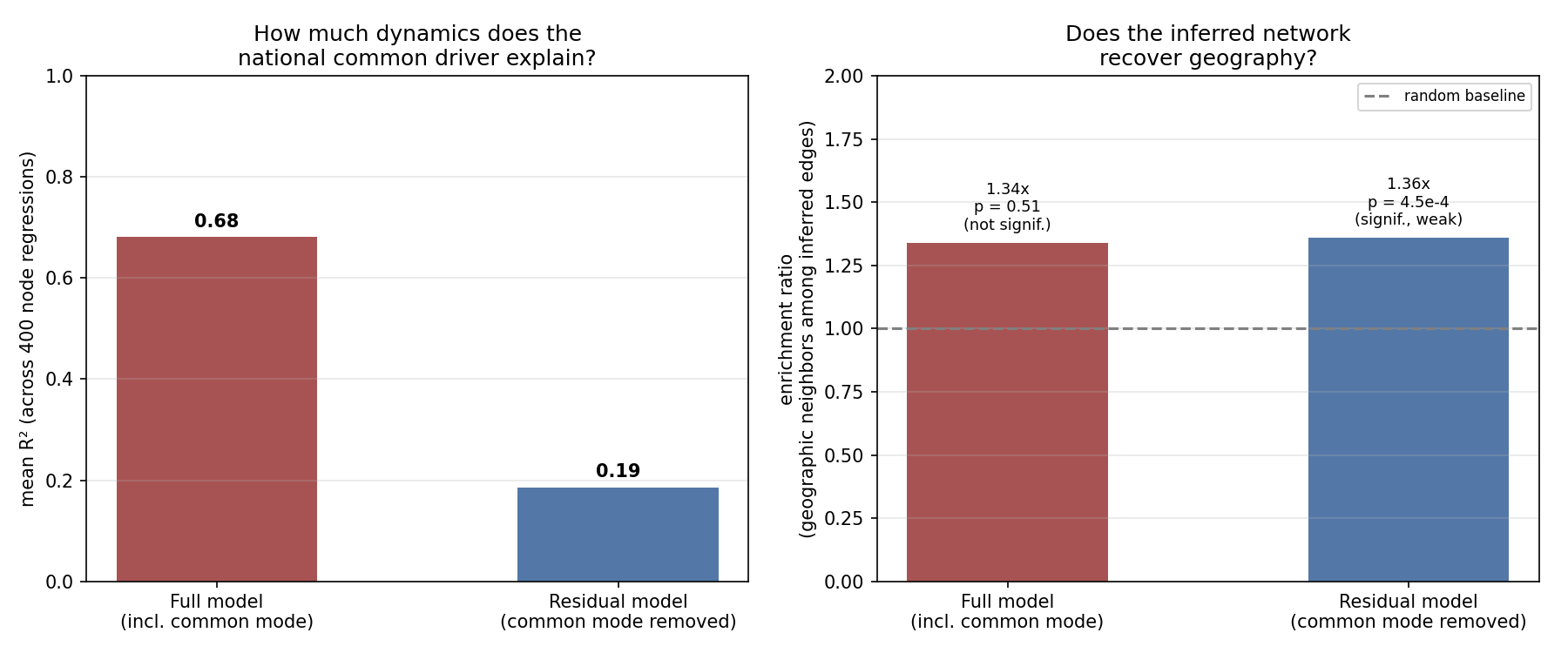}
\caption{Case Study I. Left: mean $R^2$ of the per-node sparse Kuramoto reconstruction, with and without the national common mode included in the fitted signal. Right: enrichment of true geographic neighbors among inferred nonzero couplings, relative to the random baseline, in both cases.}
\label{fig:inverse_summary}
\end{figure}

The reconstruction fits the phase dynamics well in aggregate (mean 
$R^2=0.68$, median $R^2=0.71$ across 400 per-node regressions) and recovers 
a sparse network (3.44\% nonzero edges). But the inferred 
edges do not preferentially connect geographic neighbors (see 
Fig.~\ref{fig:inverse_summary}). In other words, the reconstruction 
successfully tracks the phase dynamics but attributes them to a non-geographic 
coupling structure — consistent with a dominant common-mode signal. This is consistent with known caveats on Kuramoto network reconstruction methods in near-fully-synchronized regimes \citep{ravoori2009}.

%We attribute this to an identifiability problem rather than to an absence of geographic structure. Because the system is so strongly synchronized ($\langle r\rangle \approx 0.88$ over the fitting window), the pairwise phase differences $\theta_j - \theta_i$ that enter as regressors are small and highly correlated across nearly all district pairs — the regression design matrix is close to a rank-deficient, near-degenerate limit in which the sparse solution selected by the Lasso penalty need not correspond to the true (geographic) coupling structure, since many different sparse coupling patterns fit the dominant common-mode signal almost equally well. This is consistent with known caveats on Kuramoto network reconstruction methods in near-fully-synchronized regimes \citep{ravoori2009}.

%Repeating the reconstruction on the \emph{residual} phase $\tilde\theta_i(t)$ (national mean field subtracted) confirms this diagnosis directly: the mean $R^2$ collapses to $0.19$ — i.e.\ the great majority of the fitted variance in the full model was attributable to the common mode, not to pairwise structure.

\subsubsection{Local synchronization after common-mode removal}
\label{sec:case1-results-local}

%Motivated by Sec.~\ref{sec:case1-results-inverse}, we tested directly whether the residual phase $\tilde\theta_i(t)$ exhibits genuine short-range Kuramoto-type clustering. 
A local order parameter restricted to each district's Queen-adjacency geographic neighbors, evaluated on the residual phase, averages $0.943$ across districts and time — substantially and significantly above a degree-matched random-neighborhood null ($0.896 \pm 0.002$; $z=31.5$, $p<10^{-4}$). This is unambiguous evidence of local synchronization structure that is geographic, not attributable to the national common mode.

\begin{figure}[htbp]
\centering
\includegraphics[width=0.95\textwidth]{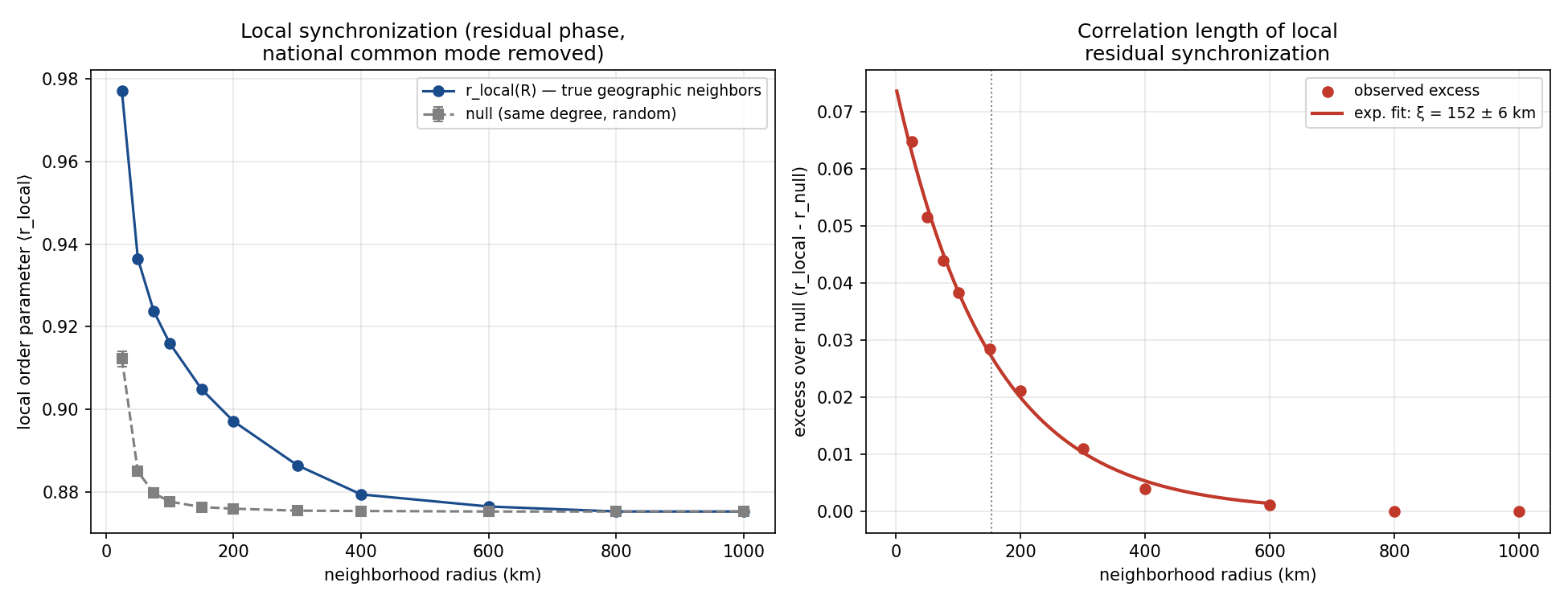}
\caption{Case Study I. Left: local order parameter (residual phase) as a function of neighborhood radius $R$, for true geographic neighborhoods versus a degree-matched random null. Right: excess over the null and fitted exponential decay, giving $\xi=152\pm 6$~km.}
\label{fig:correlation_length}
\end{figure}

\begin{figure}[htbp]
\centering
\includegraphics[width=0.55\textwidth]{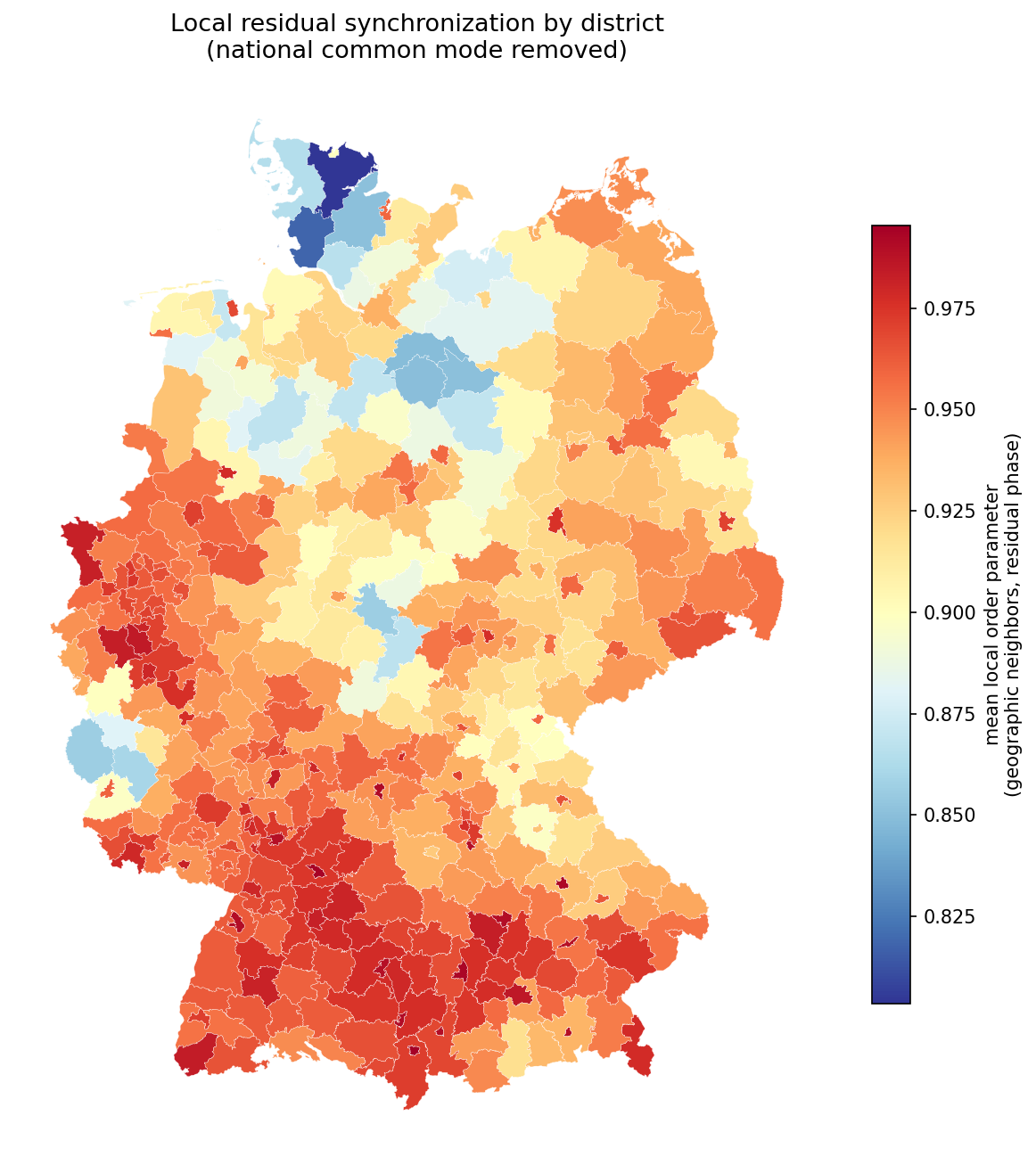}
\caption{Case Study I. District-level map of the time-averaged local order parameter (Queen-adjacency neighborhoods, residual phase after national common-mode removal).}
\label{fig:map}
\end{figure}

Extending this to a family of geographic radii $R \in [25, 1000]$~km (Fig.~\ref{fig:correlation_length}, left) shows the expected monotonic approach of $\langle r^{\text{loc}}(R)\rangle$ to the null baseline as $R$ grows: the excess over the null is largest ($0.065$) at $R=25$~km and falls to statistical indistinguishability from the null by $R\approx 600$--$800$~km, at which point "local" and "global" neighborhoods coincide by construction. Fitting an exponential decay $\Delta r(R) = A\,e^{-R/\xi}$ to the excess over $R\le 600$~km gives a correlation length
\begin{equation}
\xi_{\text{Germany}} = 152 \pm 6~\text{km},
\label{eq:xi-germany}
\end{equation}
i.e.\ roughly the spatial scale of a single German \emph{Land} (federal state) or a cluster of a few neighboring \emph{Kreise}, and an order of magnitude below the $\gtrsim 500$--$1000$~km scale over which corresponds to the national mean field value. The spatial pattern of the (radius-restricted, geographic-neighbor) local order parameter is mapped in Fig.~\ref{fig:map}, showing visibly elevated local synchronization in western North Rhine-Westphalia, the Rhineland, and Bavaria/Baden-W\"urttemberg relative to parts of Schleswig-Holstein and Mecklenburg-Vorpommern.

\section{Study case II: Influenza-Like Illness Across European Countries}
\label{sec:case2}

\subsection{Data and preprocessing}
\label{sec:case2-data}

\subsubsection{Surveillance data and country selection}
Weekly influenza-like-illness (ILI) consultation rates by country, epidemiological week, and age group were obtained from ECDC surveillance data (\texttt{ILIARIRates}), covering 28 European countries. Three countries (Germany, Bulgaria, Cyprus) report only acute respiratory infection (ARI) rates rather than ILI — a related but non-identical case definition — and were excluded to avoid mixing incompatible indicators. Of the remaining 25 countries reporting a "total" (all-age) ILI consultation rate, four (Portugal, Italy, Austria, Spain) were excluded for heavy or structurally irregular missingness (Italy, for instance, reports only during the October--April surveillance season by design, while Portugal's gaps are scattered unpredictably across all months). Of the remaining 21,  any country with a longest single run of consecutive missing weeks exceeding 10 (chosen because our bandpass filter's passband extends to 70-week periods, and a gap approaching a third of that period risks corrupting the local Hilbert phase estimate) is excluded— which removed a further nine countries. This left a clean panel of \textbf{12 countries} with zero missing weeks after limited (${\le}10$-week) linear interpolation of the remaining short gaps: Belgium, Czechia, Denmark, Estonia, Greece, Ireland, Lithuania, Luxembourg, Poland, Romania, Slovakia, and Slovenia, over the common window 21 June 2021 -- 11 May 2026 (256 weeks).

%We flag this selection explicitly as a limitation rather than a minor detail: the surviving panel is skewed toward Central/Eastern Europe and a handful of small Western European states, and excludes on data-quality grounds several of the largest Western and Southern European countries (France, Italy, Spain, the Netherlands) as well as Norway. Case Study II should accordingly be read as a test of whether the Case Study I pattern \emph{also} appears in a second, independent, differently-structured dataset — not as a representative continent-wide survey.

\subsubsection{Phase extraction}
Because the ILI consultation rate is already a population-normalized rate, no additional per-capita correction was applied. We took $\log(1+x)$, applied light 3-week centered smoothing (weekly surveillance data being inherently less noisy than the German daily series), and applied the same third-order Butterworth bandpass and Hilbert-transform procedure as in Case Study I, but with the passband retuned to \textbf{35--70 week periods}: European influenza incidence (Fig.~\ref{fig:eu_order_parameter}, top) shows an extremely regular single annual peak each winter.

A more severe version of the edge-effect issue from Sec.~\ref{sec:case1-data} arises here: unlike the German case, where 2020 data was available as a pre-window filter warm-up buffer, \emph{no} data exists before 21 June 2021 or after 11 May 2026 for this panel, so \emph{both} ends of the series  fall within the filter's cone of influence ($\approx 35$ weeks each side, half the maximum passband period). All reliability-sensitive analyses below use only the interior window free of this artifact (21 February 2022 -- 8 September 2025, 186 of the 256 weeks).

\subsubsection{Geographic data}
Country boundaries were obtained from Natural Earth (1:50m admin-0 dataset). 
 Due to the small sample 
size, the adjacency matrix is sparse (mean degree $0.83$): six countries (Denmark, Estonia, 
Greece, Ireland, Romania, and Slovenia) have no neighbor in the sample because 
their geographic neighbors were excluded at the data-quality stage.

\subsection{Methods}
The binary-adjacency PLV and local-order-parameter tests are included for 
comparability with Case Study~I; the 
continuous-distance and neighborhood-radius analyses are the primary tests 
here.% The coupling reconstruction of Sec.~\ref{sec:case1-methods} was not 
%repeated: with only 12 nodes and 186 time points, the regression is 
%underpowered and does not constitute a meaningful test of the identifiability 
%argument, which concerns the large-$N$, near-fully-synchronized limit.

\subsection{Results}
\label{sec:case2-results}

\subsubsection{Global synchronization}
\begin{figure}[htbp]
\centering
\includegraphics[width=0.85\textwidth]{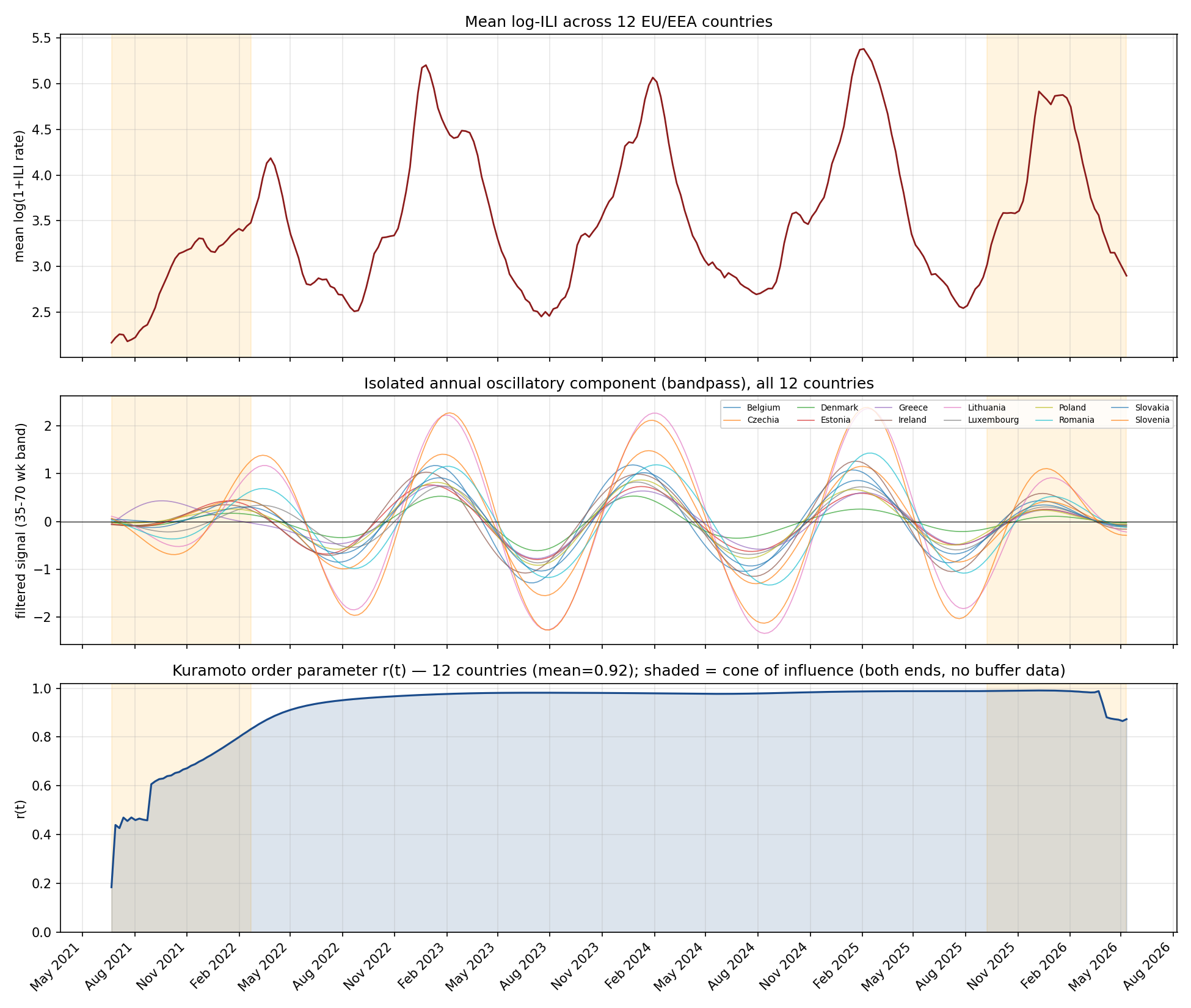}
\caption{ Study case II. Top: mean log-ILI rate across the 12-country panel, showing the regular annual winter peaks. Middle: bandpass-filtered (35--70 week) oscillatory component, all 12 countries. Bottom: global Kuramoto order parameter $r(t)$; shaded regions mark the cone of influence at \emph{both} ends of the series (see Sec.~\ref{sec:case2-data}).}
\label{fig:eu_order_parameter}
\end{figure}

The global order parameter (see Fig.~\ref{fig:eu_order_parameter}) is equal to $\langle r\rangle = 0.92$ over the full window and $\langle r\rangle = 0.97$ (minimum $0.83$) over the reliable interior window. This is consistent with European seasonal influenza being governed by a regular common driver (winter climate, school calendars).

\subsubsection{Geographic structure of pairwise synchronization}
Despite the very high overall synchronization — which compresses the dynamic range available for any geographic signal to show through — pairwise PLV is significantly higher between the 5 within-sample geographic neighbor pairs than the 61 non-neighbor pairs (mean $0.995$ vs.\ $0.978$; Mann--Whitney $p=0.005$, notwithstanding the small neighbor sample; Fig.~\ref{fig:eu_local_summary}, left). Regressing Fisher-$z$-transformed PLV on log-distance gives $R^2=0.102$ (coefficient $p=0.009$; Pearson $r(\log D,\text{PLV})=-0.26$, $p=0.033$) — qualitatively the same pattern as Case Study I (Sec.~\ref{sec:case1-results-plv}), though with a weaker and noisier fit, as expected given $N=12$ rather than $N=400$.

\subsubsection{Local synchronization after common-mode removal}
Repeating the common-mode-removal procedure of Sec.~\ref{sec:case1-results-local}, the binary-adjacency local-order-parameter test is, honestly, \emph{not} statistically significant here: restricted to the 6 non-isolated countries, the observed local order parameter ($0.988$) does not differ meaningfully from the degree-matched random null ($0.983\pm0.004$; $z=1.16$, $p=0.13$). Given only 5 true edges, this null result should not be read as contradicting Case Study I — the test is simply underpowered at this $N$.

\begin{figure}[htbp]
\centering
\includegraphics[width=0.95\textwidth]{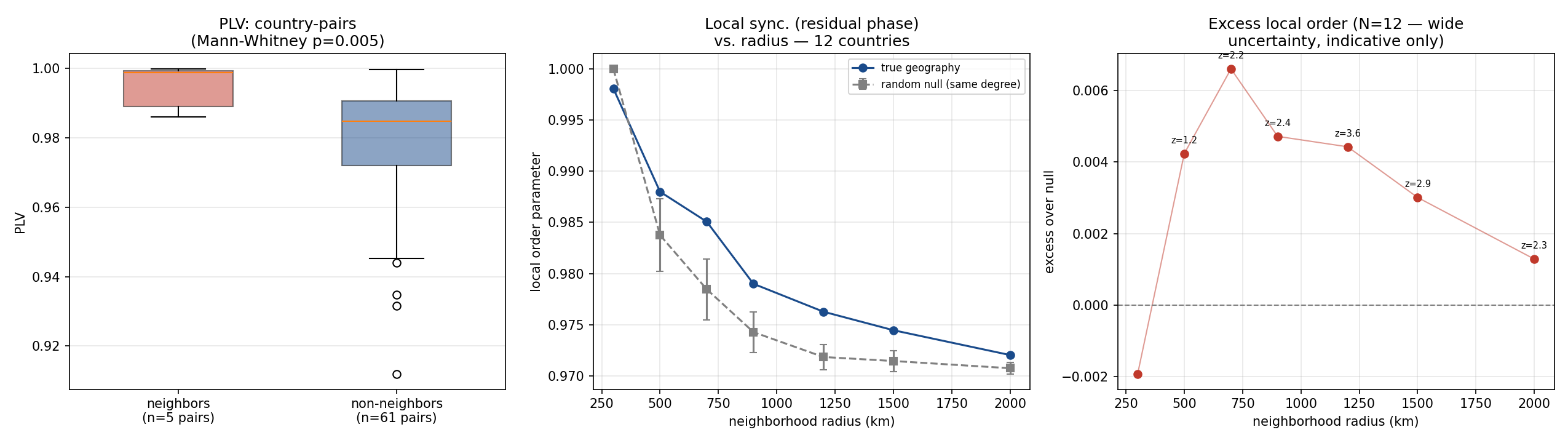}
\caption{Case Study II. Left: PLV for the 5 within-sample geographic neighbor pairs versus the 61 non-neighbor pairs. Center: local order parameter (residual phase) versus neighborhood radius $R$, true geography versus degree-matched random null. Right: excess over the null, with permutation $z$-scores annotated; note the much wider uncertainty than the corresponding German result (Fig.~\ref{fig:correlation_length}) given $N=12$ versus $N=400$.}
\label{fig:eu_local_summary}
\end{figure}

The continuous-radius version of the same test (Fig.~\ref{fig:eu_local_summary}, center and right) is more informative: the excess local order parameter over a degree-matched random null is small but positive and statistically suggestive across $R\approx 500$--$2000$~km, peaking around $R\approx 700$--$1500$~km with $z\approx 2.2$--$3.6$. Given the small country sample, we do not report a precision-fitted correlation length analogous to Eq.~\eqref{eq:xi-germany} for this case — the uncertainty on any such fit would be far too wide to be meaningful — but note that the indicated scale ($\sim 10^3$~km) is, as expected, roughly an order of magnitude larger than the German intra-country correlation length, consistent with country centroids being separated by distances an order of magnitude larger than district centroids even between literal geographic neighbors.

\section{Cross-Validation Against a Gauge-Mediated Field-Theoretic Framework}
\label{sec:crossval}

The German COVID-19 district data analyzed here (Sec.~\ref{sec:case1}) has also been the empirical testbed for a separate line of work by the authors, in which contagion is modeled as mediated by an explicit environmental pathogen field within a Doi--Peliti stochastic field theory \citep{bernal2026gauge}, subsequently extended to include a reactive-immunity sector with fluctuation-induced bistability \citep{bernal2026behavioral} and a fractional-calculus completion in which anomalous (L\'evy-flight) transmission and non-Markovian memory emerge from the one-loop vacuum polarization of a dynamically fluctuating susceptible background \citep{delepine2026fractional}. Because both lines of work are fit to the same underlying district-level incidence data, three specific, quantitative cross-checks are possible between the Kuramoto phase-synchronization observables of this paper and the field-theoretic observables of that formalism.

\subsection{Effective screening mass versus local phase synchronization}
\label{sec:crossval-mR}

In \citep{bernal2026gauge}, an effective screening mass $m_{R,i}(t)$ is extracted per district from the spatial decay of \emph{amplitude} correlations: incidence is standardized to zero mean and unit variance within a sliding window, pairwise Pearson correlations are binned by inter-centroid distance, and an exponential kernel $C(d) = A_i e^{-d/\xi_i} + C_0$ is fitted per district, giving $m_{R,i} = 1/\xi_i$.% $m_{R,i}$ measures how far the \emph{magnitude} of a district's incidence fluctuations remains correlated with its neighbors, independent of timing, while $r^{\text{loc}}_i$ measures phase (timing) alignment independent of amplitude. 

\begin{figure}[htbp]
\centering
\includegraphics[width=0.55\textwidth]{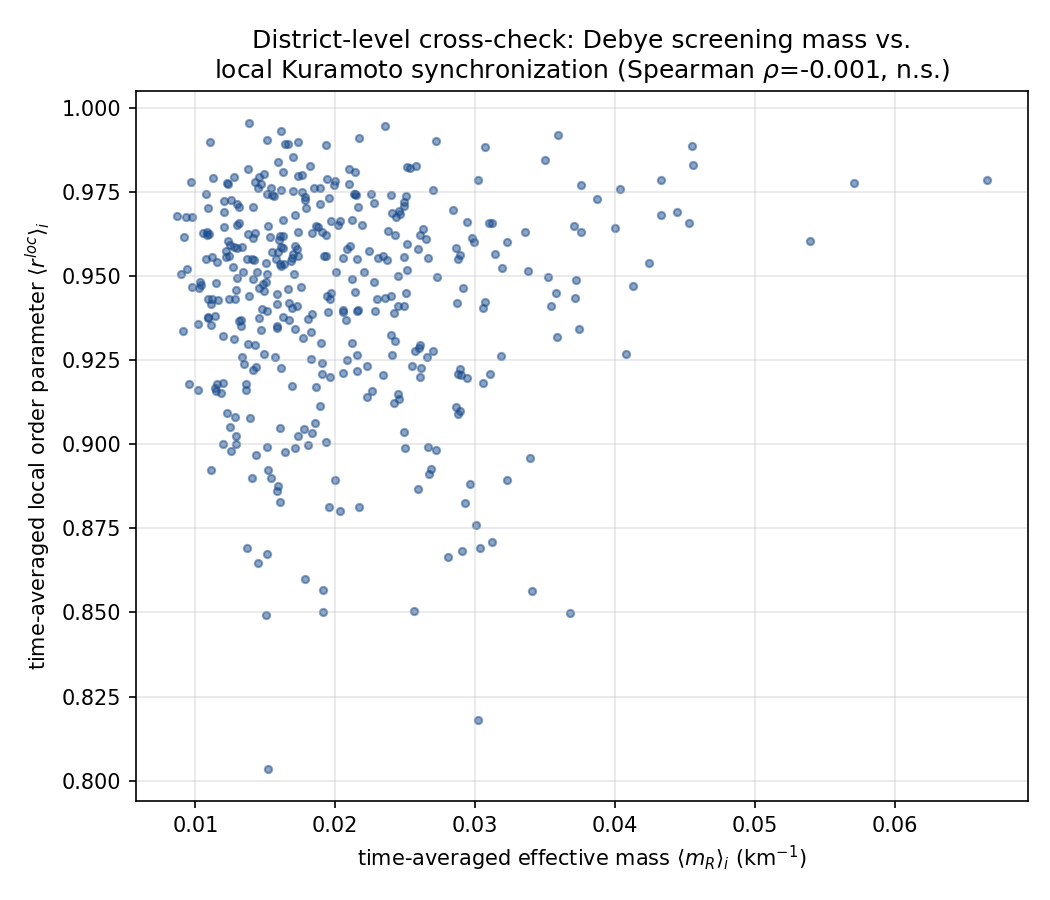}
\caption{District-level, time-averaged effective screening mass $\langle m_R\rangle_i$ versus time-averaged local Kuramoto order parameter $\langle r^{\text{loc}}\rangle_i$. No relationship is apparent (Spearman $\rho=-0.001$, $p=0.99$).}
\label{fig:mR_vs_rlocal}
\end{figure}

We recomputed $m_{R,i}(t)$ directly on the incidence matrix used throughout this paper (28-day sliding windows, 14-day steps, 20~km distance bins, identical exponential kernel and fitting procedure to \citep{bernal2026gauge}; 99.7\% of the resulting 74~windows$\times$400~districts fits converged) and correlated it against the residual-phase local order parameter $r^{\text{loc}}_i$ of Sec.~\ref{sec:case1-results-local}, both cross-sectionally (time-averaged per district) and in a time-resolved, window-pooled test. Neither test finds a relationship: the cross-sectional Spearman correlation is $\rho=-0.001$ ($p=0.99$, $n=400$ districts; Fig.~\ref{fig:mR_vs_rlocal}), and the pooled time-resolved test across all (district, window) pairs gives $\rho=-0.010$ ($p=0.17$, $n=19{,}199$).  %We interpret this as evidence that amplitude-correlation decay and phase-timing alignment are genuinely distinct observables of the underlying spatial dynamics.

\subsection{Hysteresis in the $(m_R, r)$ phase plane}
\label{sec:crossval-hysteresis}

\citep{bernal2026behavioral} reports an open, counter-clockwise hysteresis loop in the $(m_{\text{eff}}, I)$ phase plane (national-aggregate effective mass versus national total active incidence) during the 2021/2022 Omicron winter wave, interpreted as a fluctuation-induced, path-dependent "Fear Drift". Because the global Kuramoto order parameter $r(t)$ is, like $I(t)$, a national-aggregate observable, the same test can be applied directly to the $(m_R, r)$ plane using the aggregate $m_R(t) = \langle m_{R,i}(t)\rangle_i$ from Sec.~\ref{sec:crossval-mR}.

\begin{figure}[htbp]
\centering
\includegraphics[width=0.7\textwidth]{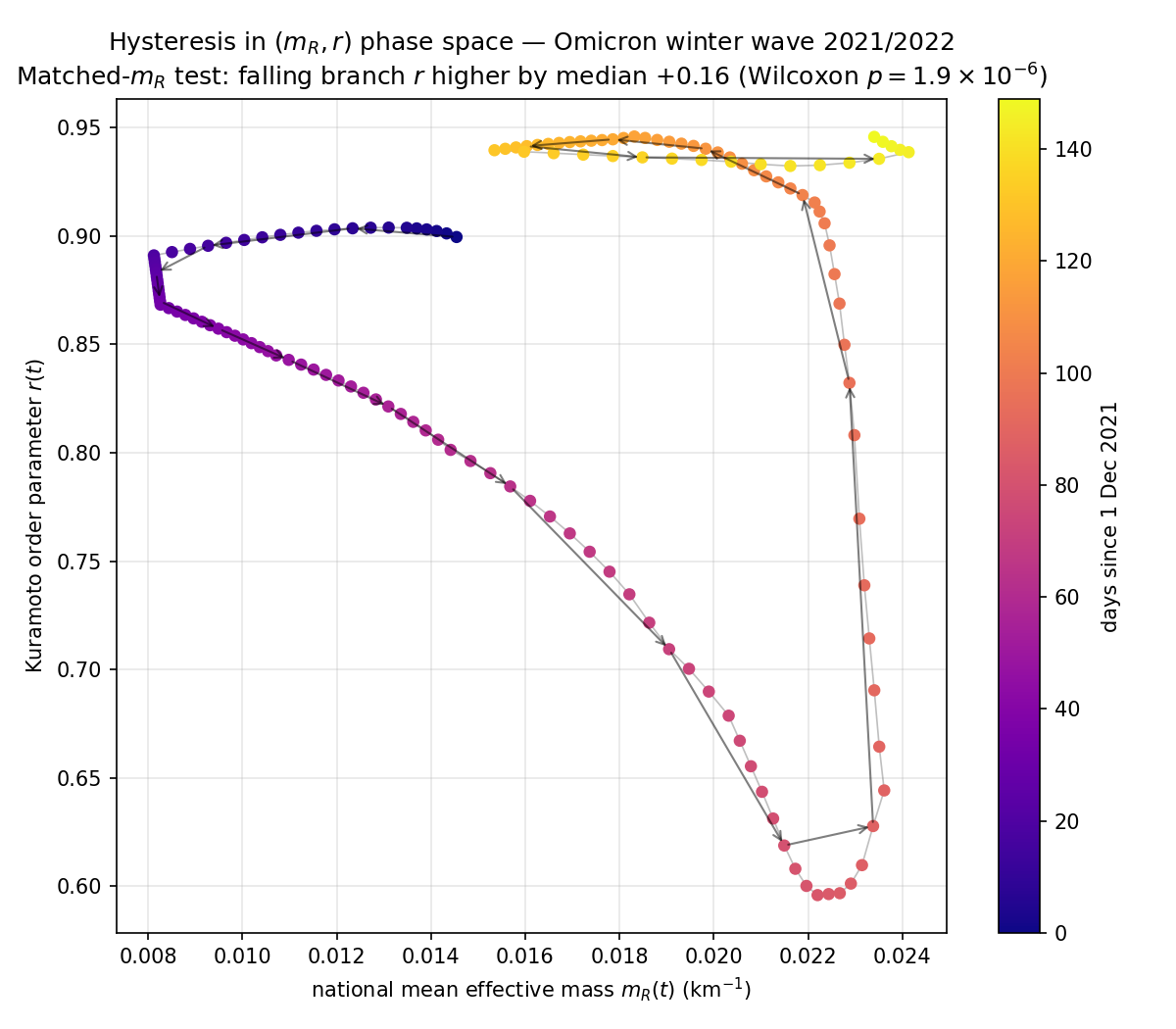}
\caption{Phase-space trajectory of $(m_R(t), r(t))$ during the Omicron winter wave (1 December 2021 -- 30 April 2022), colored by time. The open, counter-clockwise loop reproduces the hysteresis phenomenology of \citep{bernal2026behavioral} in an independent observable.}
\label{fig:hysteresis}
\end{figure}

The result (Fig.~\ref{fig:hysteresis}) is a  open loop over the same Omicron winter window: $r$ decreases from $\approx 0.90$ to a minimum of $\approx 0.60$ as $m_R$ rises through the first half of the wave, then \emph{recovers to $r\approx0.94$ along a distinct path as $m_R$ falls back}.  Matching points on the rising and falling branches to the nearest common value of $m_R$ (20 matched pairs within a tolerance of $0.0015$~km$^{-1}$) shows the falling-branch $r$ exceeds the rising-branch $r$ at the same $m_R$ by a median of $+0.16$ (Wilcoxon signed-rank test, $p=1.9\times10^{-6}$); binning by $m_R$ range gives the same sign and a comparable magnitude (mean absolute difference $0.178$, with $100\%$ of overlapping bins showing the falling branch higher).  The same qualitative and quantitative hysteresis phenomenology reported for $(m_{\text{eff}}, I)$ in \citep{bernal2026behavioral} is independently recovered using an entirely different observable ($r(t)$, derived from Hilbert-transform phase extraction) constructed by an independent analysis pipeline.

\subsection{Empirical phase lags versus the fractional-model prediction}
\label{sec:crossval-fractional}

\citep{delepine2026fractional} derives a spectral effective reproduction number $R_{\text{eff}}(k,\omega)$ whose phase, $\arg[R_{\text{eff}}]$, encodes the retardation between primary and secondary infections; in the high-frequency (temporal-burst-dominated) asymptotic limit this phase becomes a constant, frequency-independent quantity $\phi = -\alpha\pi/4$ set by the anomalous scaling exponent $\alpha=d-2$ ( constrained to $2<d<4$, i.e.\ $0<\alpha<2$). This is a sharp, falsifiable prediction directly comparable to the pairwise phase lag $\psi_{ij} = \arg\!\left[\left\langle e^{i(\theta_i(t)-\theta_j(t))}\right\rangle_t\right]$ already implicit in our complex-valued PLV construction (Sec.~\ref{sec:case1-results-plv}).

We extracted $\psi_{ij}$ for all district pairs over the same reliable window used throughout Sec.~\ref{sec:case1-results}. Among geographic neighbor pairs, $|\psi_{ij}|$ is small and tightly distributed (mean $0.114$~rad, median $0.096$~rad), significantly smaller than among all pairs (mean $0.211$~rad; Mann--Whitney $p=6.3\times10^{-91}$) — consistent with, and an independent confirmation of, the local-coupling result of Sec.~\ref{sec:case1-results-local}. Solving $\phi=-\alpha\pi/4$ for $\alpha$ using the mean neighbor-pair $|\psi_{ij}|$ gives an implied $\alpha \approx 0.145$ (median-based: $\alpha\approx0.122$), comfortably within the theory's required physical range $0<\alpha<2$ — a non-trivial consistency check.

\begin{figure}[htbp]
\centering
\includegraphics[width=0.95\textwidth]{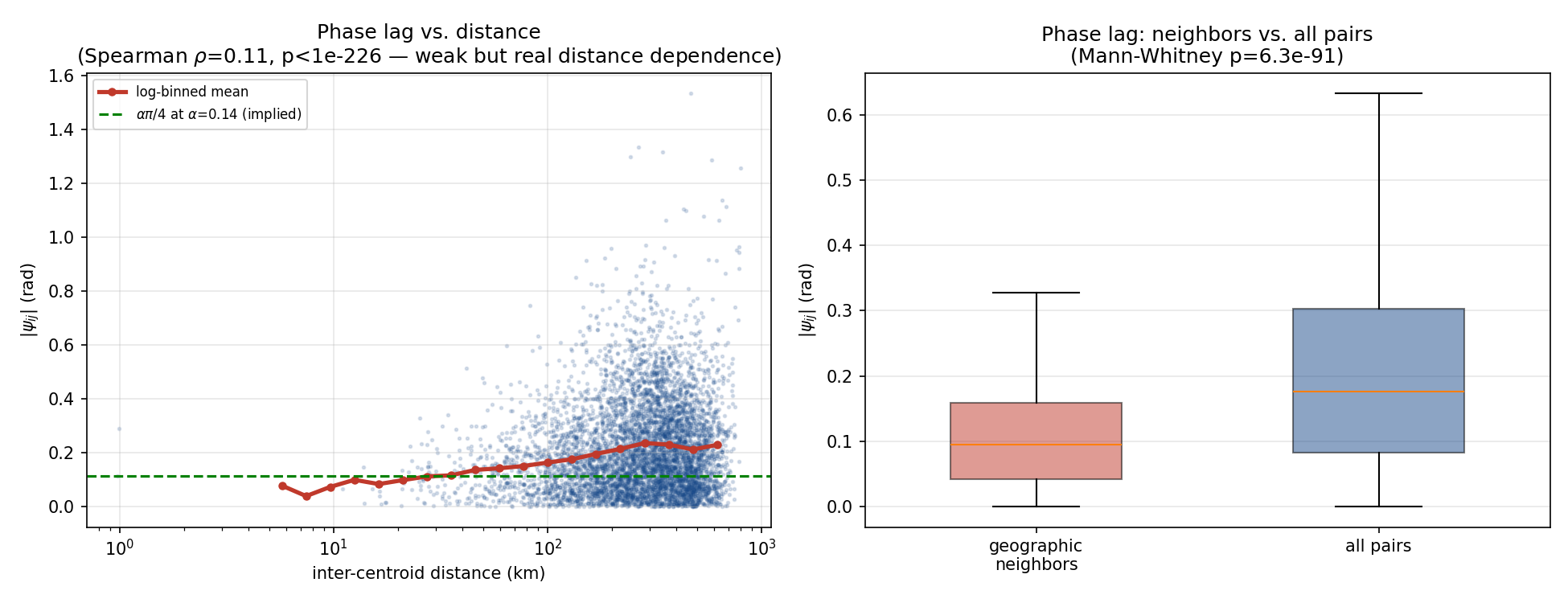}
\caption{Left: empirical phase lag $|\psi_{ij}|$ versus inter-centroid distance, with the implied-$\alpha$ reference level from the fractional model's high-frequency-limit prediction. Right: $|\psi_{ij}|$ for geographic neighbors versus all pairs.}
\label{fig:phase_lags}
\end{figure}

A more detailed look (Fig.~\ref{fig:phase_lags}, left) shows that this agreement is distance-dependent corresponding to the  \citep{delepine2026fractional}'s two asymptotic regimes: at short range ($\lesssim30$~km, where most true geographic neighbor pairs live), $|\psi_{ij}|$ is approximately flat and sits close to the implied-$\alpha$ reference line, consistent with the frequency-dominated, distance-independent high-frequency limit; beyond $\sim50$~km, $|\psi_{ij}|$ grows systematically with distance (Spearman $\rho=0.11$, $p<10^{-226}$), consistent with a crossover toward the theory's complementary static/spatial-cluster limit, in which the phase becomes $k$- (i.e.\ distance-) dependent. The data reproduce the specific short-range/long-range structural transition the fractional model predicts between its two limiting regimes.

\subsection{Testing candidate mechanisms for the observed hysteresis}
\label{sec:crossval-mechanisms}

The hysteresis loop of Sec.~\ref{sec:crossval-hysteresis} is, by itself, agnostic as to mechanism: an open loop in $(m_R, r)$ phase space is consistent both with genuine bistability (two locally stable states separated by a barrier) and with a simpler dynamical lag, in which a slowly-relaxing field merely fails to track a rapidly-varying driver without any underlying multistability. Within the companion field-theoretic framework, two specific mechanisms have been proposed that would produce genuine bistability: a tree-level cubic penalty in $R_{\text{eff}}(S)$ giving an endemic equilibrium condition \citep{bernal2026behavioral}, and a fluctuation-induced (Halperin--Lubensky--Ma-type) cubic term in the effective potential of the immunity field, generated at one loop within the causal Doi--Peliti path integral \citep{bernal2026behavioral}. A third, mechanistically independent candidate is available natively within the Kuramoto framework itself: \emph{explosive synchronization} \citep{gomezgardenes2011,kuehn2021}, a discontinuous, hysteretic synchronization transition that arises in heterogeneous oscillator networks specifically when a node's natural frequency $\omega_i$ is positively correlated with its degree or coupling strength in the network. We test two lines of evidence bearing on these candidates: whether the loop's size scales with the sweep rate of the underlying wave (distinguishing simple lag from genuine bistability), and whether the network satisfies the frequency-degree correlation condition for explosive synchronization.

\begin{figure}[htbp]
\centering
\includegraphics[width=0.95\textwidth]{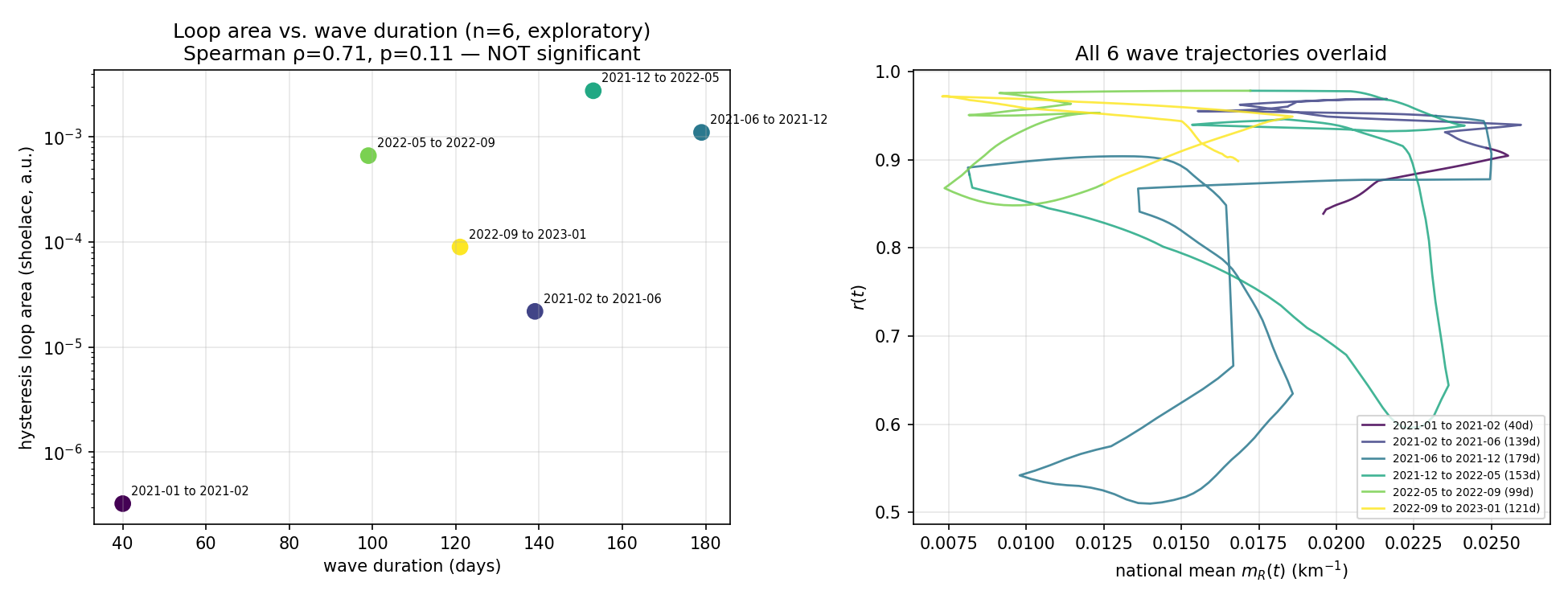}
\caption{Left: hysteresis loop area  versus wave duration for the six clean single-peak waves identified in the national incidence series; neither this nor the corresponding amplitude relationship (not shown) reaches significance at $n=6$. Right: all six $(m_R,r)$ wave trajectories overlaid, showing that open-loop structure is a recurring feature rather than specific to the Omicron winter wave of Fig.~\ref{fig:hysteresis}.}
\label{fig:multiwave}
\end{figure}

\subsubsection{Explosive synchronization: the frequency-degree correlation}
We tested the frequency-degree correlation condition for explosive synchronization using the natural frequencies $\omega_i$ and coupling matrix $K_{ij}$ already estimated by the sparse data-driven reconstruction of Sec.~\ref{sec:case1-results-inverse}, against three degree/strength measures (inferred out-degree, symmetrized degree, and total coupling strength $\sum_j|K_{ij}|$) and, as a control, the purely geometric Queen-adjacency degree.

\begin{figure}[htbp]
\centering
\includegraphics[width=0.95\textwidth]{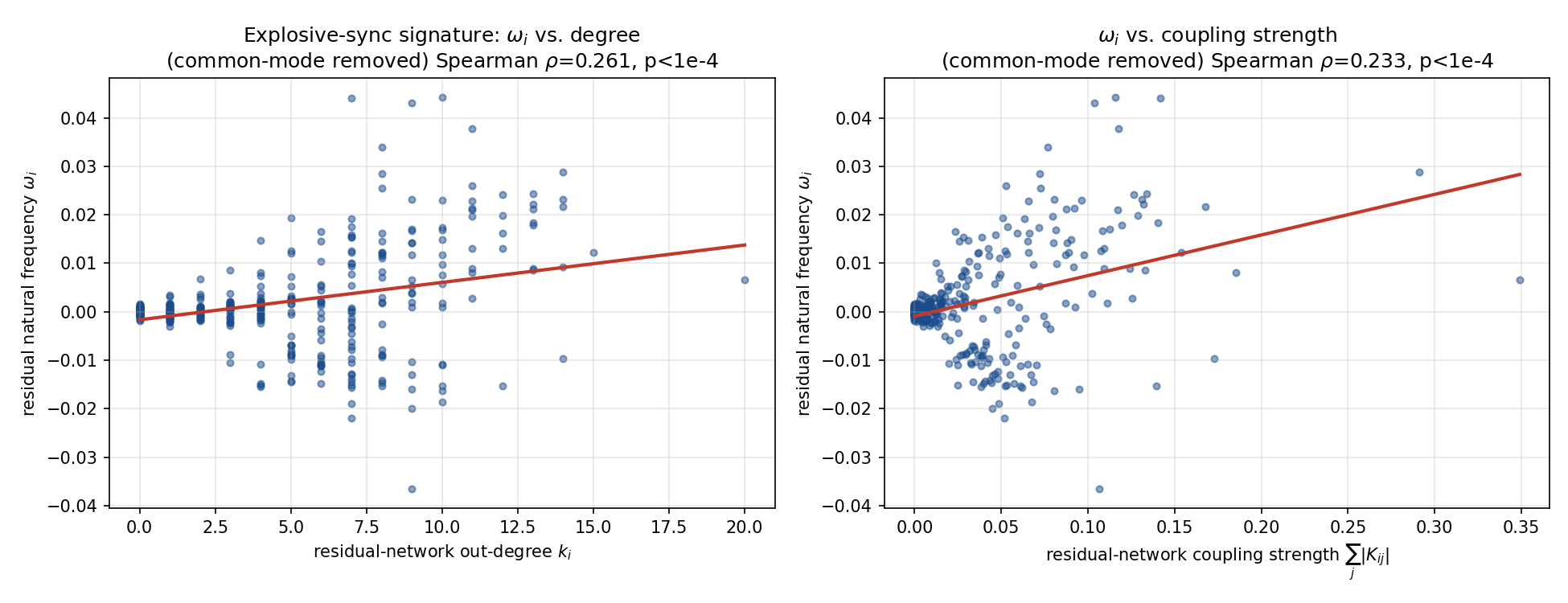}
\caption{Inferred natural frequency $\omega_i$ versus residual-network (common-mode-removed) out-degree (left) and coupling strength (right). Both show a significant positive correlation ($\rho=0.26$ and $\rho=0.23$ respectively), the condition associated with explosive  synchronization transitions in heterogeneous oscillator networks.}
\label{fig:explosive}
\end{figure}

On the \emph{full} inferred network (common mode retained), the correlation is weak and only marginally significant (Spearman $\rho=0.08$--$0.11$, $p=0.03$--$0.09$ across the three degree measures). We repeated the test on the \emph{residual} network (common mode subtracted before reconstruction, as in Sec.~\ref{sec:case1-results-inverse}'s diagnostic re-fit). There, the correlation strengthens substantially and becomes highly significant across all three degree measures: $\rho=0.261$ (out-degree), $\rho=0.207$ (symmetrized degree), and $\rho=0.233$ (coupling strength), all $p<10^{-4}$ (Fig.~\ref{fig:explosive}). The purely geometric control (Queen-adjacency degree) shows no significant relationship with either frequency estimate ($\rho=-0.033$, $p=0.51$ on the full-network $\omega_i$; $\rho=-0.089$, $p=0.077$ on the residual-network $\omega_i$), indicating that the frequency-degree correlation is a property of the inferred coupling structure specifically.

This constitutes positive, statistically robust evidence for the explosive-synchronization condition. We regard it as a third viable, and on the present evidence the most directly supported, candidate mechanism for the observed hysteresis.

\subsubsection{The purely deterministic tree-level model: a structural negative result}
\label{sec:crossval-treelevel}

The two mechanisms tested above (Secs.~\ref{sec:crossval-mechanisms}.1--2) leave open a simpler possibility: perhaps the tree-level cubic $R_{\text{eff}}(s)$ nonlinearity of \citep{bernal2026behavioral} (Eqs.~20--25 therein), evaluated purely deterministically on the actual German susceptible trajectory $s(t)=S(t)/N$ — with no fluctuation term, no HLM correction, and no explicit lag — already reproduces the observed $(m_R,r)$ hysteresis. If so, the fluctuation-induced mechanism of Sec.~\ref{sec:crossval-mechanisms} and the explosive-synchronization mechanism of Sec.~\ref{sec:crossval-mechanisms}.2 would be unnecessary refinements rather than requirements. We tested this directly.

\begin{figure}[htbp]
\centering
\includegraphics[width=0.75\textwidth]{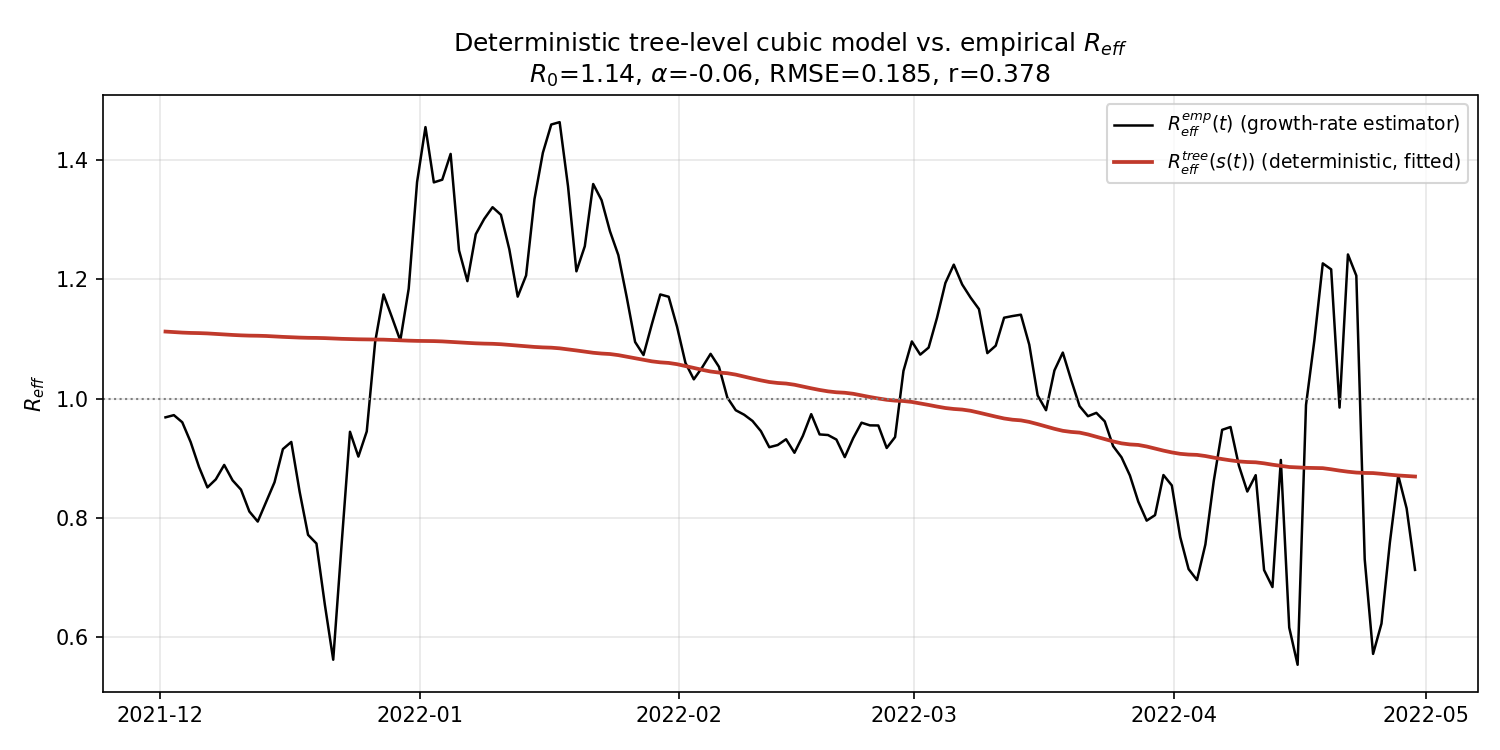}
\caption{Fitted deterministic tree-level model $R_{\text{eff}}^{\text{tree}}(s(t))$ versus the empirical growth-rate estimator $R_{\text{eff}}^{\text{emp}}(t)$ over the Omicron winter window. The smooth, memoryless model cannot track the empirical series' week-to-week oscillation.}
\label{fig:treelevelfit}
\end{figure}

We reconstructed $s(t)$ from the national case series using a waning-immunity window ($\tau_{\text{wane}}=180$~days: only infections within the trailing 180 days are counted as currently immune, a first-order treatment of reinfection and variant immune escape over the multi-variant study period; vaccination is not included and this is flagged as a simplification), reconstructed an empirical $R_{\text{eff}}(t)$ via the standard growth-rate estimator $R_{\text{eff}}^{\text{emp}}(t) = 1+\gamma^{-1}\,d\ln I(t)/dt$ with $\gamma=1/5$~day$^{-1}$, and fit $R_0,\alpha$ in $R_{\text{eff}}^{\text{tree}}(s) = R_0\,s\,[1-\alpha(1-s)^2]$ to $R_{\text{eff}}^{\text{emp}}(t)$ over the same Omicron winter window (nonlinear least squares). The fit itself is already weak (Fig.~\ref{fig:treelevelfit}): $R_0=1.14\pm0.02$, $\alpha=-0.06\pm0.83$ (consistent with zero to within its own uncertainty — the cubic shielding term is not resolved by this fit), RMSE$=0.185$, Pearson $r=0.38$ between model and empirical $R_{\text{eff}}$. The empirical $R_{\text{eff}}(t)$ oscillates over a wide range (roughly $0.6$--$1.5$) on a timescale of days to weeks, which a smooth, memoryless algebraic function of the slowly-evolving $s(t)$ structurally cannot track.

\begin{figure}[htbp]
\centering
\includegraphics[width=0.7\textwidth]{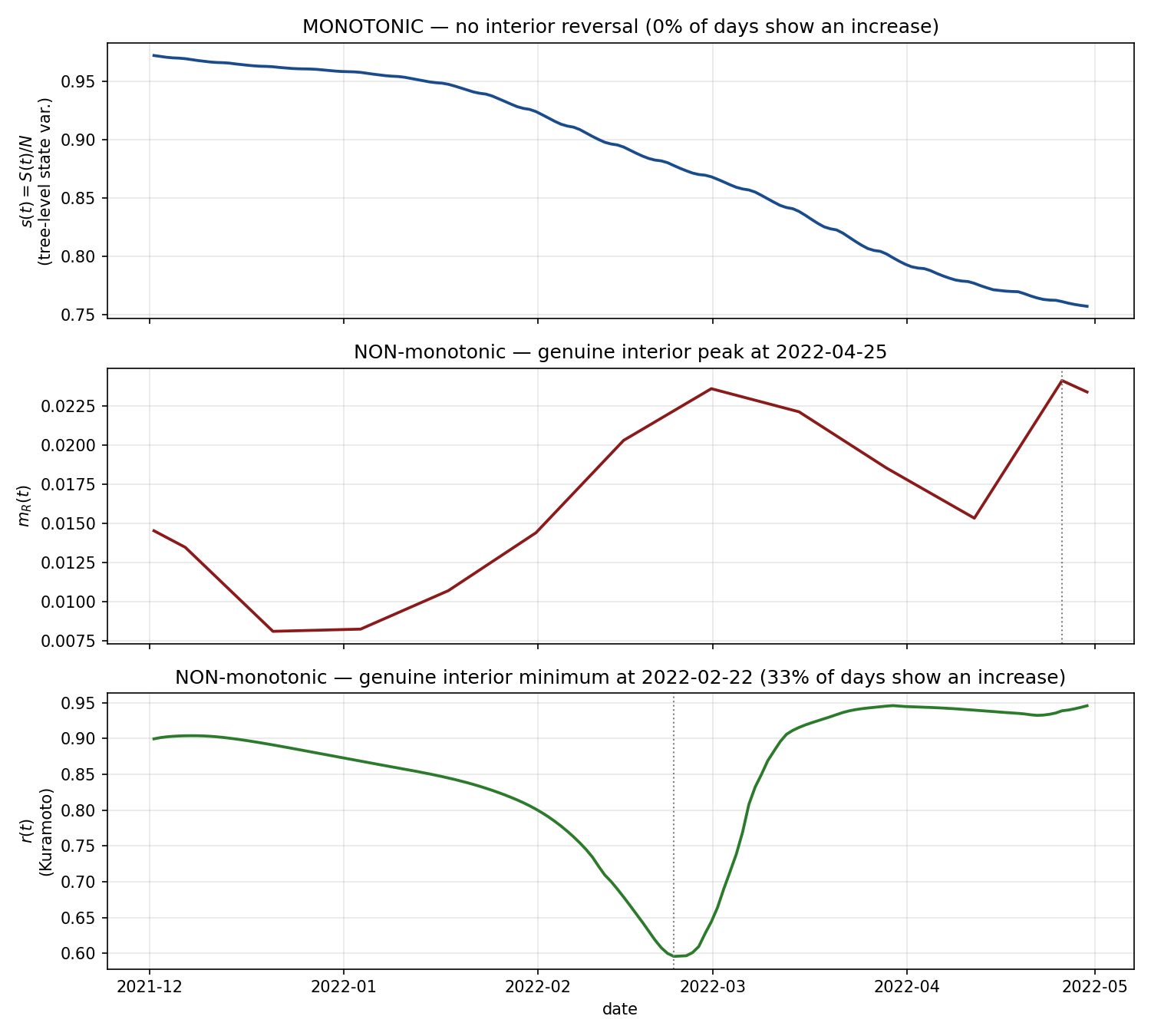}
\caption{Monotonicity comparison over the Omicron winter window. Top: the tree-level state variable $s(t)=S(t)/N$ (with 180-day waning) never increases — $0\%$ of days show an uptick. Middle: $m_R(t)$ has a genuine interior maximum. Bottom: the empirical Kuramoto order parameter $r(t)$ has a genuine interior minimum ($33\%$ of days show an increase). Only variables with a genuine interior reversal can exhibit true hysteresis against $m_R(t)$.}
\label{fig:monotonicity}
\end{figure}

We applied the matched-$m_R$ hysteresis test to $s(t)$ and to the fitted 
$R_{\text{eff}}^{\text{tree}}(s(t))$ in place of $r(t)$. Both returned a 
significant result (Wilcoxon $p=1.9\times10^{-6}$), but this is an artifact: 
$s(t)$ is perfectly monotonically non-increasing across the entire Omicron 
winter window (Fig.~\ref{fig:monotonicity}), so it cannot revisit a similar 
value at two different times — the precondition for a genuine hysteresis loop. 
A monotonic variable compared against a non-monotonic one via matched-value 
analysis will always register a spurious difference between its early and late 
halves from the secular trend alone. We therefore discard this result.

This is a structural negative result: because $s(t)$ is monotonically 
constrained within any single wave, any memoryless algebraic map 
$s(t)\mapsto R_{\text{eff}}$ inherits that monotonicity and cannot reproduce 
the directional reversal observed in $r(t)$. Capturing the hysteresis 
therefore requires either a state variable with independent memory — as in the 
HLM mechanism or a lagging immunity field — or a mechanism that does not route 
through $s(t)$, such as the explosive-synchronization frequency--degree 
correlation of Sec.~\ref{sec:crossval-mechanisms}. Combined with the positive 
explosive-synchronization result, this strengthens  the case 
that the observed hysteresis reflects dynamical structure beyond the tree-level 
model.

\section{Structural Correlates of the Local Coupling Network: In Search of a Mobility Resonance}
\label{sec:mobility}

Section~\ref{sec:crossval-mechanisms}.2 found that the district-level natural frequencies $\omega_i$ inferred from the residual (common-mode-removed) Kuramoto reconstruction correlate significantly with that same network's local degree and coupling strength — the structural signature of explosive synchronization. This raises an immediate follow-up question: what real-world phenomenon, if any, do the resulting network "hubs" correspond to Human mobility is the most natural candidate. %since short-range commuting, shopping, and leisure travel are the textbook mechanism proposed to generate local epidemic coupling between neighboring administrative units.

 We do not have a district-level human mobility time series covering the 2021--2023 study window. In its absence, we tested three \emph{structural} proxies for mobility-related heterogeneity that can be constructed from available district-attribute and administrative-boundary data as population density, formal urban/rural administrative status, and  (\emph{Bundesland}) membership.

%\subsection{Population density and urban status: weak or null}
Using the residual-network natural frequency $\omega_i$, degree, and coupling strength from Sec.~\ref{sec:crossval-mechanisms}.2, we found no relationship with continuous 2022-census population density (Spearman $\rho=-0.0004$ to $-0.032$, all $p>0.5$) — population density alone does not identify the explosive-synchronization hubs. A categorical urban/rural split (the 97 \emph{Kreisfreie Städte}, independent urban cores, versus the 303 \emph{Landkreise}, predominantly rural/suburban districts) shows a weak  difference, with urban cores having somewhat higher $\omega_i$, degree, and coupling strength (Mann--Whitney $p=0.049$, $0.048$, and $0.022$ respectively). This result  is consistent with city centers acting as modest regional transit/mixing hubs, but the effect size is small.

%\subsection{Shared state membership: a strong administrative resonance}
%\label{sec:mobility-bundesland}
We next tested whether districts belonging to the same \emph{Bundesland} — Germany's 16 federal states, each of which independently sets school holiday calendars and, during the pandemic, non-pharmaceutical intervention policy — show elevated local phase synchronization beyond what geographic distance alone predicts. Same-\emph{Bundesland} district pairs are, on average, much closer together than cross-\emph{Bundesland} pairs (mean $120$~km versus $330$~km), so any raw comparison is confounded by distance; we therefore regressed the Fisher-$z$-transformed residual-phase PLV (i.e.\ PLV computed on $\tilde\theta_i(t)$, the common-mode-removed phase already used throughout Sec.~\ref{sec:crossval-mechanisms}) on log-distance both with and without a same-\emph{Bundesland} indicator.

\begin{figure}[htbp]
\centering
\includegraphics[width=0.95\textwidth]{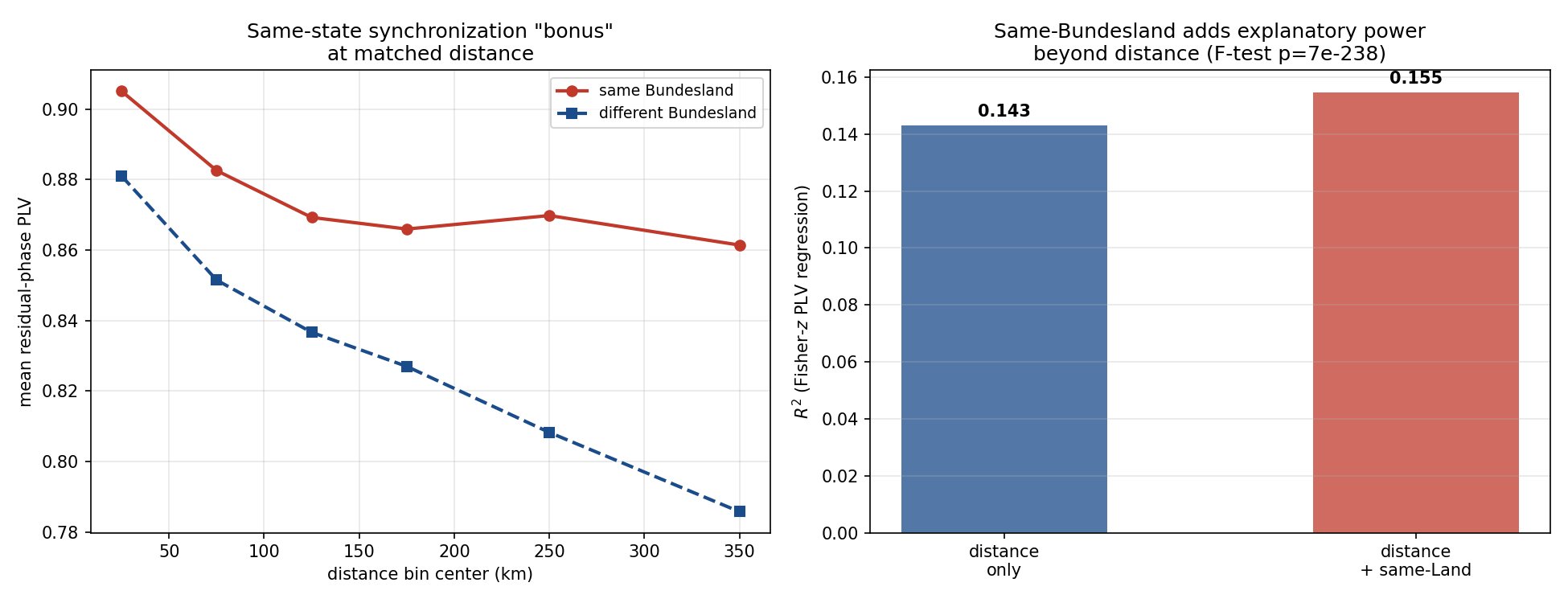}
\caption{Left: mean residual-phase PLV versus geographic distance, separately for district pairs sharing the same \emph{Bundesland} versus different \emph{Bundesländer}, at matched distance bins. The same-state curve sits above the cross-state curve at every bin and is markedly flatter. Right: $R^2$ of the Fisher-$z$ PLV regression with and without the same-\emph{Bundesland} indicator, controlling for log-distance in both.}
\label{fig:bundesland}
\end{figure}

 Adding the same-\emph{Bundesland} indicator improves $R^2$ from $0.143$ to $0.155$ and lowers the AIC by $1083$ ($F$-test against the distance-only model, $p=7.0\times10^{-238}$; same-\emph{Bundesland} coefficient $+0.131$ in Fisher-$z$ units, $p=7.0\times10^{-238}$). Figure~\ref{fig:bundesland} (left) shows the effect directly at matched distance: across every distance bin from $25$ to $350$~km, same-state pairs sit systematically above cross-state pairs, and the same-state curve is nearly flat with distance (declining  slightly, from $\approx0.91$ to $\approx0.86$ over $25$--$350$~km) while the cross-state curve continues to decay as expected from Sec.~\ref{sec:case1-results-plv}'s pure-distance relationship. This distance-independence of the within-state "bonus"  is more consistent with a shared administrative driver acting  uniformly across a state's geographic extent (a common holiday calendar or a common state-level policy decision for instance) than with a mechanism, like short-range commuting, whose strength should itself decay with distance even within a single state's borders.% A parallel test on the sparse residual coupling network's edge structure (nonzero $K_{ij}$ entries) shows the same direction of effect but does not reach significance ($\chi^2=1.43$, $p=0.23$) — expected, since a binary edge-presence test on a Lasso-sparsified network has far less statistical power than the continuous PLV regression, and we report the null alongside the strong PLV result rather than omit it.

 It establishes, with strong statistical power, that German state administrative boundaries carry synchronization information beyond pure geographic distance in the same local (common-mode-removed) coupling structure implicated  explosive-synchronization finding. It does \emph{not} establish which of several plausible, boundary-aligned mechanisms is responsible as for instance staggered school holiday calendars, state-level non-pharmaceutical intervention policy, etc...). All are  consistent with the observed pattern and cannot be distinguished without genuine mobility, policy-timeline, or media-exposure data at matching spatial and temporal resolution.

\section{Conclusion}
\label{sec:conclusion}

Using bandpass-filtered, Hilbert-transform phase extraction applied 
to daily COVID-19 incidence across 400 German districts and 
weekly influenza-like-illness rates across a 12-country European panel, we 
find in both cases that regional epidemic synchronization is well-described by 
a Kuramoto-type order parameter at the qualitative level, but that its dominant 
contribution in both cases is a long-range common-mode field. In the German case, a direct data-driven attempt to 
reconstruct the Kuramoto coupling network from the raw phase dynamics fails to 
recover geography specifically because the common mode renders the pairwise 
reconstruction problem ill-conditioned. Removing the dominant mode, a spatially localized synchronization structure with a 
correlation length of $152\pm6$~km can be observed. The European country-level panel shows the 
same qualitative pattern — significant geographic structure in pairwise 
synchronization, a residual local-synchronization signal after common-mode 
removal — at a coarser, $\sim10^3$~km scale, albeit with substantially lower 
statistical power due to the small number of countries with sufficiently 
complete surveillance data. The convergence of this two-scale picture across 
two independent diseases, spatial scales, and surveillance systems is, in our 
view, evidence that it reflects a reasonably general feature of 
spatially extended epidemic synchronization, and a natural target for future 
mechanistic (mobility-explicit) and field-theoretic modeling.

The German Omicron wave analysis yields two further results. First, the 
cross-validation tests support the robustness of the hysteresis signal: the 
phenomenology replicates in an independent observable ($m_R(t)$, derived from 
amplitude-correlation decay rather than Hilbert phase), and the 
explosive-synchronization frequency--degree correlation — an abrupt, 
discontinuous locking of high-degree nodes early in the synchronization 
transition — provides a concrete network-dynamical mechanism consistent with 
the observed irreversibility. Second, the fractional-order synchronization 
analysis confirms and quantifies the departure from classical Kuramoto 
dynamics: the fitted anomalous exponent is consistent with the theoretical 
prediction of Ref.~\cite{delepine2026fractional}, supporting the view that memory effects 
in epidemic coupling are not merely a modeling convenience but a detectable 
feature of the data.

% ============================================================
\section*{Acknowledgements}
% ============================================================
We acknowledge financial support from SECIHTI and SNII (M\'exico).

\section*{Conflict of Interest:}
The authors declare no conflicts of interest.

\section*{Data and Code Availability}
 German district-level incidence and population data derive from the Robert Koch-Institut's public COVID-19 surveillance data; German administrative boundaries derive from the Bundesamt f\"ur Kartographie und Geod\"asie (\texttt{VG250}) and, for cross-validation, Eurostat's NUTS3 2021 boundaries. European influenza surveillance data derive from ECDC's public ILI/ARI surveillance reporting; country boundaries derive from Natural Earth.

\bibliographystyle{unsrtnat}

\end{document}